\documentclass[twocolumn,english,aps,prl,superscriptaddress,address,showpacs,showacknowledgments,longbibliography]{revtex4-1}
\pdfoutput=1
\usepackage[latin9]{inputenc}
\usepackage{babel}
\usepackage{float}
\usepackage{amsmath}
\usepackage{amssymb}
\usepackage{graphicx}
\usepackage{esint}
\usepackage[unicode=true,pdfusetitle,
 bookmarks=true,bookmarksnumbered=false,bookmarksopen=false,
 breaklinks=false,pdfborder={0 0 1},backref=false,colorlinks=false]
 {hyperref}

\makeatletter
\usepackage{babel}
\usepackage{babel}
\usepackage{babel}
\usepackage{babel}
\usepackage{bbold}
\hypersetup{
   pdftitle={},
   pdfauthor={},
   colorlinks=true,       % false: boxed links; true: colored links
   linkcolor=red,          % color of internal links (change box color with linkbordercolor)
   citecolor=green,        % color of links to bibliography
   filecolor=magenta,      % color of file links
   urlcolor=blue           % color of external links
}


\makeatother

\begin{document}

\title{From Kardar-Parisi-Zhang scaling to explosive desynchronization in
arrays of limit-cycle oscillators}

\author{Roland Lauter}

\affiliation{Institut für Theoretische Physik II, Friedrich-Alexander-Universität
Erlangen-Nürnberg, Staudtstr. 7, 91058 Erlangen, Germany}

\affiliation{Max Planck Institute for the Science of Light, Günther-Scharowsky-Straße
1/Bau 24, 91058 Erlangen, Germany}

\author{Aditi Mitra}

\affiliation{Department of Physics, New York University, 4 Washington Place, New
York, NY 10003, USA}

\author{Florian Marquardt}

\affiliation{Institut für Theoretische Physik II, Friedrich-Alexander-Universität
Erlangen-Nürnberg, Staudtstr. 7, 91058 Erlangen, Germany}

\affiliation{Max Planck Institute for the Science of Light, Günther-Scharowsky-Straße
1/Bau 24, 91058 Erlangen, Germany}
\begin{abstract}
We study the synchronization physics of 1D and 2D oscillator lattices
subject to noise and predict a dynamical transition that leads to
a sudden drastic increase of phase diffusion. Our analysis is based
on the widely applicable Kuramoto-Sakaguchi model, with local couplings
between oscillators. For smooth phase fields, the time evolution can
initially be described by a surface growth model, the Kardar-Parisi-Zhang
(KPZ) theory. We delineate the regime in which one can indeed observe
the universal KPZ scaling in 1D lattices. For larger couplings, both
in 1D and 2D, we observe a stochastic dynamical instability that is
linked to an apparent finite-time singularity in a related KPZ lattice
model. This has direct consequences for the frequency stability of
coupled oscillator lattices, and it precludes the observation of non-Gaussian
KPZ-scaling in 2D lattices.
\end{abstract}
\maketitle
Networks and lattices of coupled limit-cycle oscillators do not only
represent a paradigmatic system in nonlinear dynamics, but are also
highly relevant for potential applications. This significance derives
from the fact that the coupling can serve to counteract the effects
of the noise that is unavoidable in real physical systems. Synchronization
between oscillators can drastically suppress the diffusion of the
oscillation phases, improving the overall frequency stability. Experimental
implementations of coupled oscillators include laser arrays \textbf{\cite{2013_Soriano_Garcia-Ojalvo_Mirasso_Fischer_Complex_Photonics_Dynamics_and_Applications}}
and coupled electromagnetic circuits, e.g.~\cite{2015_English_Zheng_Mertens_Experimental_Study_of_Synchronization_of_Coupled_Electrical_Self-oscillators,2012_Temirbayev_Zhanabaev_Tarasov_Ponomarenko_Rosenblum_Experiments_on_oscillator_ensembles_with_global_nonlinear_coupling},
as well as the modern recent example of coupled electromechanical
and optomechanical oscillators \cite{Lipson_Sync_using_Light,Zhang_Lipson_Synced_array_PRL,Tang_Sync_racetrack,Seshia_Sync_micromechanical_oscillators,Roukes_Sync_nanomechanical_oscillators}.
In this work, we will be dealing with the experimentally most relevant
case of locally coupled 1D and 2D lattices.

Naive arguments indicate that the diffusion rate of the collective
phase in a coupled lattice of $N$ synchronized oscillators is suppressed
as $1/N$, which leads to the improvement of frequency stability mentioned
above. However, it is far from guaranteed that this ideal limit is
reached in practice \cite{Cross_Improving_Frequency_Precision,Allen_Cross_Frequency_precision_with_spirals}.
The nonequilibrium nonlinear stochastic dynamics of the underlying
lattice field theory is sufficiently complex that a more detailed
analysis is called for. In this context, it has been conjectured earlier
that there is a fruitful connection \cite{Synchronization_universal_concept}
between the synchronization dynamics of a noisy oscillator lattice
and the Kardar-Parisi-Zhang (KPZ) theory of stochastic surface growth
\cite{KPZ_original,Halpin-Healy_Zhang_review}. 

We have been able to confirm that this is indeed true in a limited
regime, particularly for 1D lattices. However, the most important
prediction of our analysis consists in the observation that a certain
dynamical instability can take the lattice system out of this regime
in the course of the time evolution. As we will show, this instability
is related to an apparent finite-time singularity in the evolution
of the related KPZ lattice model. It has a significant impact on the
phase dynamics, increasing the phase spread by several orders of magnitude.
As such, this phenomenon represents an important general feature of
the dynamics of coupled oscillator lattices.

\begin{figure}[t]
\centering{}\includegraphics[width=1\columnwidth]{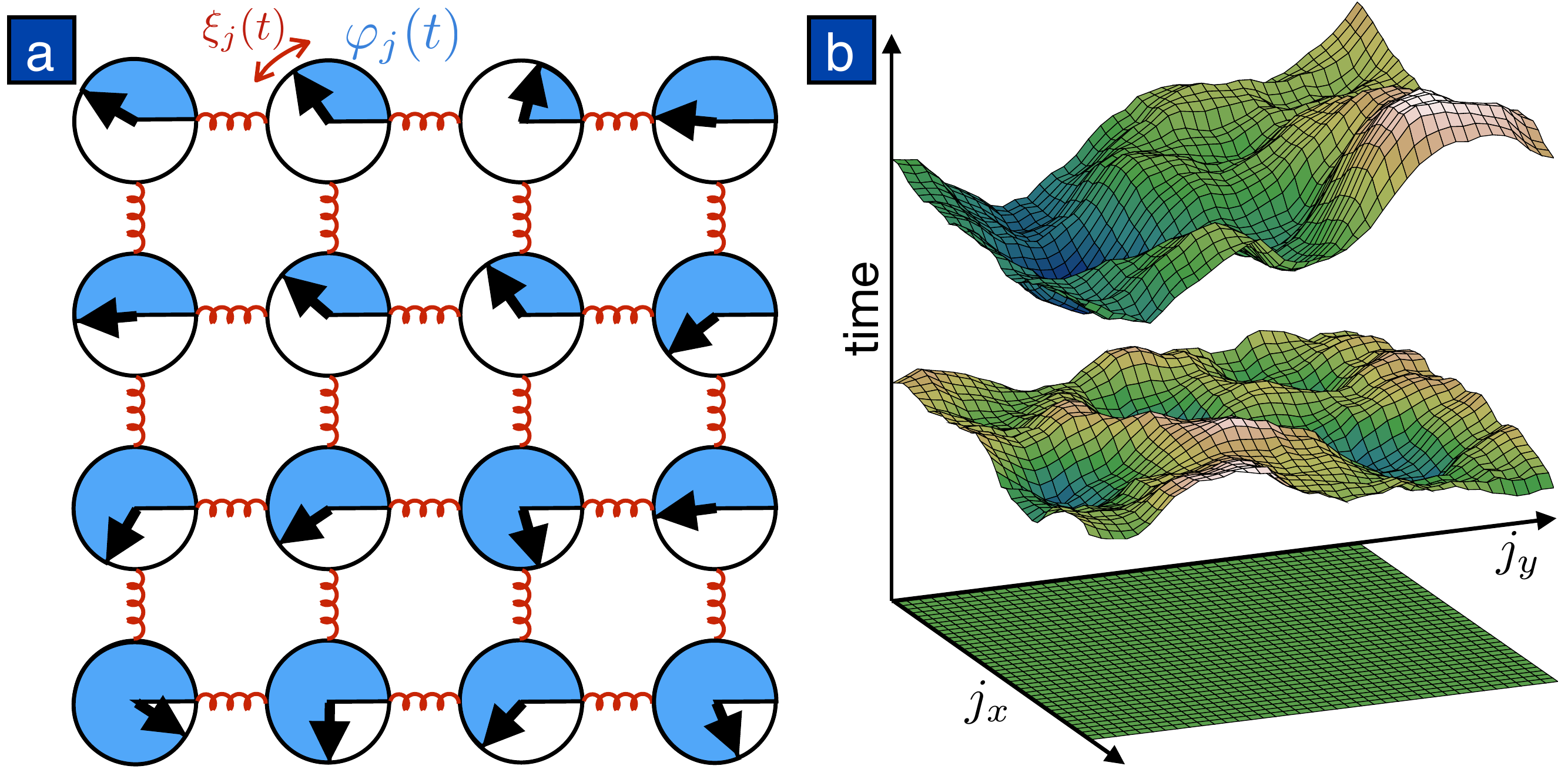}\protect\caption{(a) Scheme of an oscillator array. We consider one- and two-dimensional
arrays of limit-cycle oscillators, which are described individually
by their phases $\varphi_{j}(t)$. These phases are influenced by
white Gaussian noise $\xi_{j}(t)$ and coupling to their nearest neighbors,
see Eq.~(\ref{eq:Kuramoto-Sakaguchi_model}). (b) Stochastic time
evolution of the phase field in a 2D array of coupled oscillators
(smoothed for clarity). The field is flat initially and roughens with
time.\label{fig:Model_scheme}}
\end{figure}

Before we turn to a definition of the model, it is helpful to briefly
outline the wider context of this study. We will be dealing with phase-only
models, which can often be used to describe systems of coupled limit-cycle
oscillators effectively (see also Fig.~\ref{fig:Model_scheme}),
whenever the amplitude degree of freedom is irrelevant. The most prominent
examples are the Kuramoto model and extensions thereof \cite{Kuramoto_original,Kuramoto_Lattice_of_Rings,Sakaguchi_Kuramoto_Rotater_Model,Acebron_Spigler_Review}.
These deterministic models are studied intensely for their synchronization
properties \cite{Synchronization_universal_concept}, mainly for globally
coupled systems with disorder, as well as for pattern formation (for
locally coupled systems) \cite{Cross_Pattern_Formation_review}. In
the latter case, interesting effects show up even in the absence of
disorder \cite{Sakaguchi01051988}. 

Adding noise to these models can significantly influence the synchronization
properties \cite{Acebron_Spigler_Review}; see \cite{1991_Strogatz_Mirollo_Stability_of_Incoherence}
for an example in globally coupled systems. In locally coupled systems,
the delicate interplay between the nonlinear coupling, the noise,
and the spatial patterns can lead to even more complex dynamics. In
contrast to the related XY model \cite{Kosterlitz_Thouless_original},
which is used to describe systems in thermodynamic equilibrium, driven
nonlinear oscillator lattices are usually far from equilibrium. This
is reflected in an additional, ``non-variational'' coupling term
\cite{Kuramoto_Lattice_of_Rings,Sakaguchi_Kuramoto_Rotater_Model}.
As long as the phase field is smooth, one can employ a continuum description
of the oscillator lattice \cite{Synchronization_universal_concept}.
This is important to make the connection to the theory of surface
growth. The continuum description also links our research to recent
developments in the study of (non-equilibrium) driven dissipative
condensates \cite{2016_Keeling_Sieberer_Alman_Chen_Diehl_Toner_Superfuidity_and_Phase_Correlations,2015_Sieberer_Buchhold_Diehl_Keldysh_field_theory},
where the connection to the physics of surface growth was employed
\cite{He_Diehl_2015,2015_Ji_Gladilin_Wouters_Temporal_Coherence_of_quantum_fluids,2015_Altman_Sieberer_Chen_Diehl_Toner_Two-dimensional_Superfluidity_and_anisotropy}.
In contrast to our work, these works focus on continuum systems, where
the essential new dynamical phenomenon identified would be absent.

We start our analysis by briefly motivating the model that we will
use. Consider a large number of self-sustained nonlinear oscillators
moving on their respective limit cycles. When they are coupled, they
influence each other's amplitude and phase dynamics. Provided that
the amplitudes do not deviate much from the limit cycle, effective
models can be derived, which describe the phase dynamics only \cite{Kuramoto_original,Kuramoto_Lattice_of_Rings,Sakaguchi_Kuramoto_Rotater_Model,Acebron_Spigler_Review},
see also Fig.~\ref{fig:Model_scheme}a. The transition from the microscopic
model to the phase model depends on the physical system, with recent
examples including electromechanical \cite{Lifshitz_Collective_Dynamics_inBook}
and optomechanical \cite{Heinrich_Marquardt_Collective_Dynamics,Ludwig_Marquardt_Quantum_Many_Body}
oscillators. In real physical systems, there will be noise acting
on the oscillators. This can be modeled by a Langevin noise term in
the right hand side of the effective phase equation. 

We will be interested in understanding the competition between noise
and coupling in the context of synchronization dynamics. To bring
out fully this competition, we focus on the ideal case of a non-disordered
lattice, with uniform natural frequencies. In that case, the derivation
of an effective model leads to the noisy Kuramoto-Sakaguchi model
\cite{Sakaguchi_Kuramoto_Rotater_Model,Acebron_Spigler_Review} for
the oscillator phases $\varphi_{j}(t)$: 
\begin{align}
\dot{\varphi}_{j}= & \,S\sum_{\langle k,j\rangle}\sin(\varphi_{k}-\varphi_{j})+C\sum_{\langle k,j\rangle}\cos(\varphi_{k}-\varphi_{j})+\xi_{j},\label{eq:Kuramoto-Sakaguchi_model}
\end{align}
where $\xi_{j}(t)$ is a Gaussian white noise term with correlator
$\langle\xi_{j}(t)\xi_{k}(0)\rangle=2D_{\varphi}\delta(t)\delta_{jk}$,
and $S$ and $C$ are the coupling parameters. The sums run over nearest
neighbors. We will often call this model the ``phase model'', for
brevity. In this article, we focus on the time evolution of the phase
field from a homogeneous initial state. Hence, we set $\varphi_{j}(0)=0$
on all sites. As an illustrative example, we show the (smoothed) snapshots
of the phase field from a simulation of Eq.~(\ref{eq:Kuramoto-Sakaguchi_model})
in Fig.~\ref{fig:Model_scheme}b. In this figure, the color (and
the mesh geometry) encode the phase value at lattice site $j=(j_{x},j_{y})$
for three different points in time.

At this point it should be noted that a lattice of coupled self-sustained
oscillators actually gives rise to an additional term in the effective
phase equation. This term couples the phases of three adjacent sites.
This was derived in \cite{Heinrich_Marquardt_Collective_Dynamics,Ludwig_Marquardt_Quantum_Many_Body},
and the resulting dynamics of that extended model (without noise)
has been explored by us in a previous work \cite{Lauter_Brendel_Habraken_Marquardt_Pattern_Formation}.
However, the additional term is irrelevant for long-wavelength dynamics
and moreover can be tuned experimentally. Therefore, in the present
article we focus entirely on the important limiting case of the Kuramoto-Sakaguchi
model, Eq.~(\ref{eq:Kuramoto-Sakaguchi_model}).

How does the interplay of noise and coupling affect the frequency
stability of the oscillators? This is a central question for synchronization
and metrology. It can be discussed in terms of the \textit{average
frequencies}\textit{\emph{, defined as}}\textit{ }$\Omega_{j}(t)=t^{-1}\int_{0}^{t}\mathrm{d}t^{\prime}\dot{\varphi}_{j}(t^{\prime})=\varphi_{j}(t)/t$.
Here the $\varphi_{j}(t)$ are the \textit{phases accumulated }\textit{\emph{during
the full time evolution}} (see also \cite{Sakaguchi_PTP77_1005,Acebron_Spigler_Review}).
As the definition of $\Omega_{j}$ shows, they have an important physical
meaning in the present setting, essentially indicating the number
of cycles that have elapsed. This is in contrast to other physical
scenarios that also involve phase models like the one shown here.
For example, in studies of superfluids, the phase is defined only
up to multiples of $2\pi$. Then, the total number of phase windings
during the time evolution does not have any direct physical significance.
This distinction is important when trying to make the connections
we are going to point out below.

Important insights can be obtained from studying the evolving spread
of the average frequencies. This turns out to be directly related
to the spread of the phase field, $w_{\varphi}(t)$,
\begin{align}
w_{\varphi}^{2}(t)=\, & \big\langle\frac{1}{N}\sum_{j=1}^{N}\big[\varphi_{j}(t)-\bar{\varphi}(t)\big]^{2}\big\rangle\nonumber \\
=\, & t^{2}\big\langle\frac{1}{N}\sum_{j=1}^{N}\big[\Omega_{j}(t)-\bar{\Omega}(t)\big]^{2}\big\rangle,\label{eq:w_phi_sq}
\end{align}
where $\bar{\varphi}(t)=N^{-1}\sum_{j=1}^{N}\varphi_{j}(t)$ is the
mean (spatially averaged) phase and $\bar{\Omega}(t)=\bar{\varphi}/t$
is the mean average frequency of a lattice with $N$ sites. The angular
brackets denote an ensemble average over different realizations of
the noise.

For the simple case of uncoupled identical oscillators subject to
noise, the phase spread grows diffusively, $w_{\varphi}(t)=\sqrt{2D_{\varphi}t}$.
Hence, the spread of time-averaged frequencies decreases as $t^{-1/2}$.
This reflects the fact that the averaged frequencies are identical
in the long-time limit because there is no disorder. In this sense,
the oscillators are always synchronized. However, if coupling is included,
we will find different exponents in the time-dependency of the phase
field spread. For example, a smaller exponent means that the tendency
towards synchronization is stronger. Hence, we will see that the coupling
between the oscillators can either enhance or hinder the synchronization
process, depending on the parameter regime. We expect that this translates
to systems with small disorder in the natural frequencies.

Much of our discussion of the initial stages of evolution will hinge
on the approximations that become possible when the phases on neighboring
sites are close. Then the phase model, Eq.~(\ref{eq:Kuramoto-Sakaguchi_model}),
is well approximated by a second-order expansion in the phase differences
\cite{Synchronization_universal_concept}. This expansion can be recast
in dimensionless form using a ${\it single}$ parameter $g_{\mathrm{1d,2d}}=4D_{\varphi}C^{2}/S^{3}$.
In a one-dimensional array, for example, the resulting model reads
\begin{align}
\frac{\partial h_{j}}{\partial\tau}= & (h_{j+1}+h_{j-1}-2h_{j})\label{eq:KPZ-2}\\
 & +\frac{1}{4}[(h_{j+1}-h_{j})^{2}+(h_{j-1}-h_{j})^{2}]+\sqrt{g_{\mathrm{1d}}}\eta_{j},\nonumber 
\end{align}
where we have rescaled both the time, $\tau=St$, and the phase field,
$h_{j}=-(2C/S)(\varphi_{j}-2Ct)$. The noise correlator is $\langle\eta_{j}(\tau)\eta_{k}(0)\rangle=2\delta_{jk}\delta(\tau)$.
The generalization to two dimensions is straightforward. 

Eq.~(\ref{eq:KPZ-2}) can be readily identified as a lattice version
of the Kardar-Parisi-Zhang (KPZ) model \cite{KPZ_original,Halpin-Healy_Zhang_review,2015_Halpin-Healy_A_KPZ_Cocktail},
a universal model for surface growth and other phenomena. This nonlinear
stochastic continuum field theory describes the evolution of a height
field $h(\vec{r},t)$, 
\begin{align}
\dot{h}= & \,\nu\Delta h+\frac{\lambda}{2}(\nabla h)^{2}+\eta,\label{eq:KPZ_model}
\end{align}
with white noise $\eta(\vec{r},t)$, where $\langle\eta(\vec{r}_{1},t)\eta(\vec{r}_{2},0)\rangle=2D\delta^{d}(\vec{r}_{2}-\vec{r}_{1})\delta(t)$.
The diffusive term tries to smooth the surface, while both the noise
and the nonlinear gradient term tend to induce a roughening.

The relation of the KPZ model to coupled oscillator lattices has been
pointed out before \cite{Synchronization_universal_concept}. However,
up to now it has remained unclear how far this formal connection is
really able to predict universal features of the synchronization dynamics.
In the present article, we will indeed observe transient behavior
where universal KPZ dynamics is applicable, but we will also find
that this is invariably followed by phenomena that lead into completely
different dynamical regimes. All the numerical results discussed in
this article will refer either to the full phase model, Eq.~(\ref{eq:Kuramoto-Sakaguchi_model}),
or to its approximate version, the ``lattice KPZ model'' Eq.~(\ref{eq:KPZ-2}).
From the comparison of these models, we will be able to extract valuable
predictions for the synchronization dynamics.

It is straightforward to make the connection between Eq.~(\ref{eq:KPZ-2})
and a one-dimensional lattice version of the KPZ model more precise.
Starting from Eq.~(\ref{eq:KPZ_model}), and given a lattice constant
$a$, we have to rescale time, $\tau=(\nu/a^{2})t$, and height, $h_{j}(\tau)=(\lambda/\nu)h(x,\tau)$,
and choose a particular discretization of the derivatives. Note that
in the continuum model in one dimension, it would even be possible
to get rid of all parameters by rescaling time, height and space.
In contrast, for the lattice model, we are left with the one dimensionless
parameter $g_{\mathrm{1d}}=aD\lambda^{2}/\nu^{3}$ \cite{Moser_Kertesz_Wolf_KPZ_numerics,Halpin-Healy_Zhang_review}.
This coupling constant will become important in the following.

We had derived our lattice model, Eq.~(\ref{eq:KPZ-2}), as an approximation
to the phase model, Eq.~(\ref{eq:Kuramoto-Sakaguchi_model}), with
its trigonometric coupling terms that are periodic in the phase variables.
Hence, for the evaluation of the equation of motion, the configuration
space of each phase variable may be restricted to the compact interval
$[-\pi,\pi)$. In view of the foregoing discussion, one may then see
the phase (Kuramoto-Sakaguchi) model as a ``compact KPZ model''.
This designation has indeed been proposed in a recent article \textbf{\cite{2016_Sieberer_Wachtel_Altman_Diehl_Lattice_Duality_for_compact_KPZ}}
(see also \cite{2016_Wachtel_Sieberer_Diehl_Altman_Electrodynamic_duality_and_vortex_unbinding}).

The rescaling of time and phase introduced above, for the approximate
lattice model of Eq.~(\ref{eq:KPZ-2}), can also be employed in the
full phase model, Eq.~(\ref{eq:Kuramoto-Sakaguchi_model}). Crucially,
this leads to one more dimensionless parameter, $S/C$. For example,
the sine term will be converted to $(2C/S)\sum\sin\big[(S/2C)(h_{k}-h_{j})\big]$.
This establishes that for given differences $h_{k}-h_{j}$ the approximation,
Eq.~(\ref{eq:KPZ-2}), becomes better for smaller $S/C$. For this
reason, we will focus on small values $S/C\ll1$, where substantial
findings can be expected from the connection of the phase model to
KPZ dynamics.
\begin{figure}[t]
\centering{}\includegraphics[width=1\columnwidth]{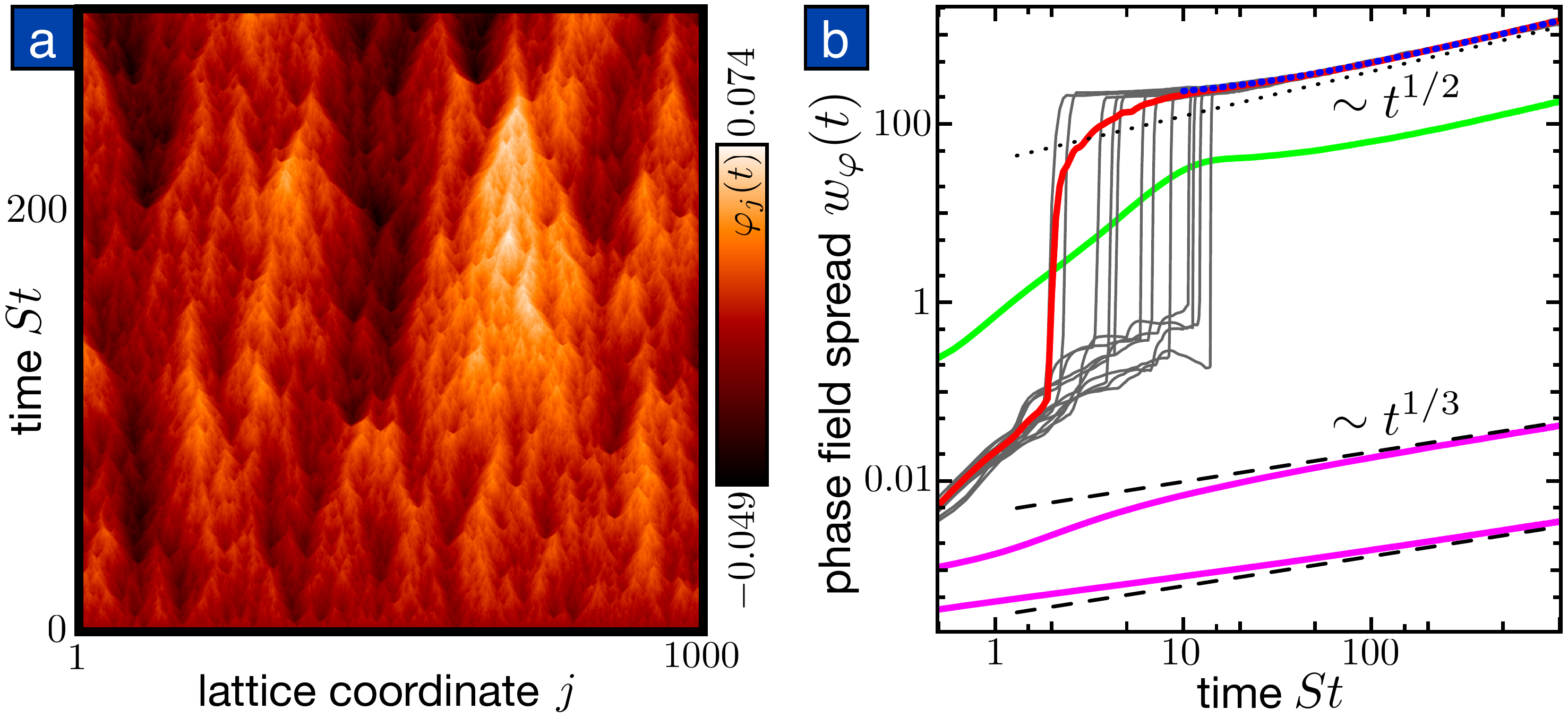}\protect\caption{Dynamics in the one-dimensional Kuramoto-Sakaguchi model, Eq.~(\ref{eq:Kuramoto-Sakaguchi_model}).
(a) Typical time evolution of the phase field from homogeneous initial
conditions. We subtracted a trivial global drift of the phases. (b)
Time evolution of the phase spread $w_{\varphi}(t)$. The magenta
curves show the simulation result for different values of the effective
coupling parameter $g_{\mathrm{1d}}=8$ (upper curve) and $g_{\mathrm{1d}}=1$
(lower curve). After an initial transient, the curves approach an
asymptotic KPZ-scaling of $w_{\varphi}(t)\propto t^{1/3}$ (dashed
black lines). For a larger value of the coupling, $g_{\mathrm{\mathrm{1d}}}=50$,
we plot examples of the phase spread from single simulations as thin
gray lines. We see a rapid increase whenever an instability occurs.
The red curve shows the small-ensemble average over 120 simulations.
After the rapid increase, it keeps growing in time, eventually approaching
diffusive behavior, $w_{\varphi}(t)\propto t^{1/2}$ (dotted black
line), which can be fitted very well by $w_{\varphi}(t)=\sqrt{A+Bt}$
(blue, dotted line). For comparison, the green curve was obtained
for another parameter set, $S/C=0.1,\ g_{\mathrm{1d}}=25$. Note the
logarithmic scale of the axes. (See the appendix for more details
on parameters.) \label{fig:1d_phase_field_width}}
\end{figure}

First insights can be gained by direct numerical simulations of the
phase model. For one-dimensional arrays, the outcome of a single simulation
is displayed in Fig.~\ref{fig:1d_phase_field_width}a. The typical
time evolution of the phase spread $w_{\varphi}(t)$ is shown in Fig.~\ref{fig:1d_phase_field_width}b.
We can distinguish two parameter regimes from the long time evolution.
In one regime, we see that after initial transients, the phase spread
evolves according to $w_{\varphi}(t)\sim t^{1/3}$ (see magenta curves).
Hence, the synchronization is enhanced as compared to the case of
uncoupled oscillators (where $w_{\varphi}(t)\sim t^{1/2}$ as discussed
above). 

The power-law growth of the phase field spread with exponent $1/3$
can be identified as universal KPZ behavior, as we will explain in
the following. Luckily, in the context of KPZ dynamics, the best-studied
quantity is the mean surface width $w$, which directly relates to
the phase spread $w_{\varphi}$ introduced above:
\begin{align}
w^{2}(L,t)= & \langle\frac{1}{L^{d}}\int\mathrm{d}^{d}r\,(h(\vec{r},t)-\bar{h}(t))^{2}\rangle,\label{eq:Squared_Surface_width}
\end{align}
with the average surface height $\bar{h}(t)=L^{-d}\int\mathrm{d}^{d}r\,h(\vec{r},t)$
in a system of linear size $L$. The surface width has been found
to obey a scaling law $w^{2}(L,t)\sim L^{2\zeta}F(t/L^{z})$ \cite{Family_Vicsek_1985}.
In particular, for $t\ll L^{z}$ (in appropriately rescaled units),
we have $w^{2}(L,t)\sim t^{2\beta}$ with $\beta=\zeta/z$. 

In one dimension, the scaling exponent can be calculated analytically
and is $\beta=1/3$ \cite{KPZ_original}. This means that the surface
will become rougher with time, but less rapidly than for independent
diffusive growth at individual sites. It is this exponent that is
also observed in the evolution of the phase model, Fig.~\ref{fig:1d_phase_field_width}.
Hence, we conclude that 1D arrays of limit-cycle oscillators, as described
by the noisy Kuramoto-Sakaguchi phase model, indeed show KPZ scaling
in certain parameter regimes.

Far more surprising is the other dynamical regime (red and green curves).
In that regime, one observes diffusive growth, $w_{\varphi}(t)\sim t^{1/2}$
for long times, which may seem unremarkable except for clearly deviating
from any KPZ predictions. However, at this point, it is worthwile
to emphasize that we are displaying curves averaged over many simulations.
If instead we look at single simulation trajectories, we see an explosive
growth of $w_{\varphi}(t)$ at some random intermediate time (gray
lines). At these random times, the phase field suddenly grows its
variance by several orders of magnitude. This corresponds to an explosive
desynchronization of the oscillators.

\begin{figure}[t]
\begin{centering}
\includegraphics[width=1\columnwidth]{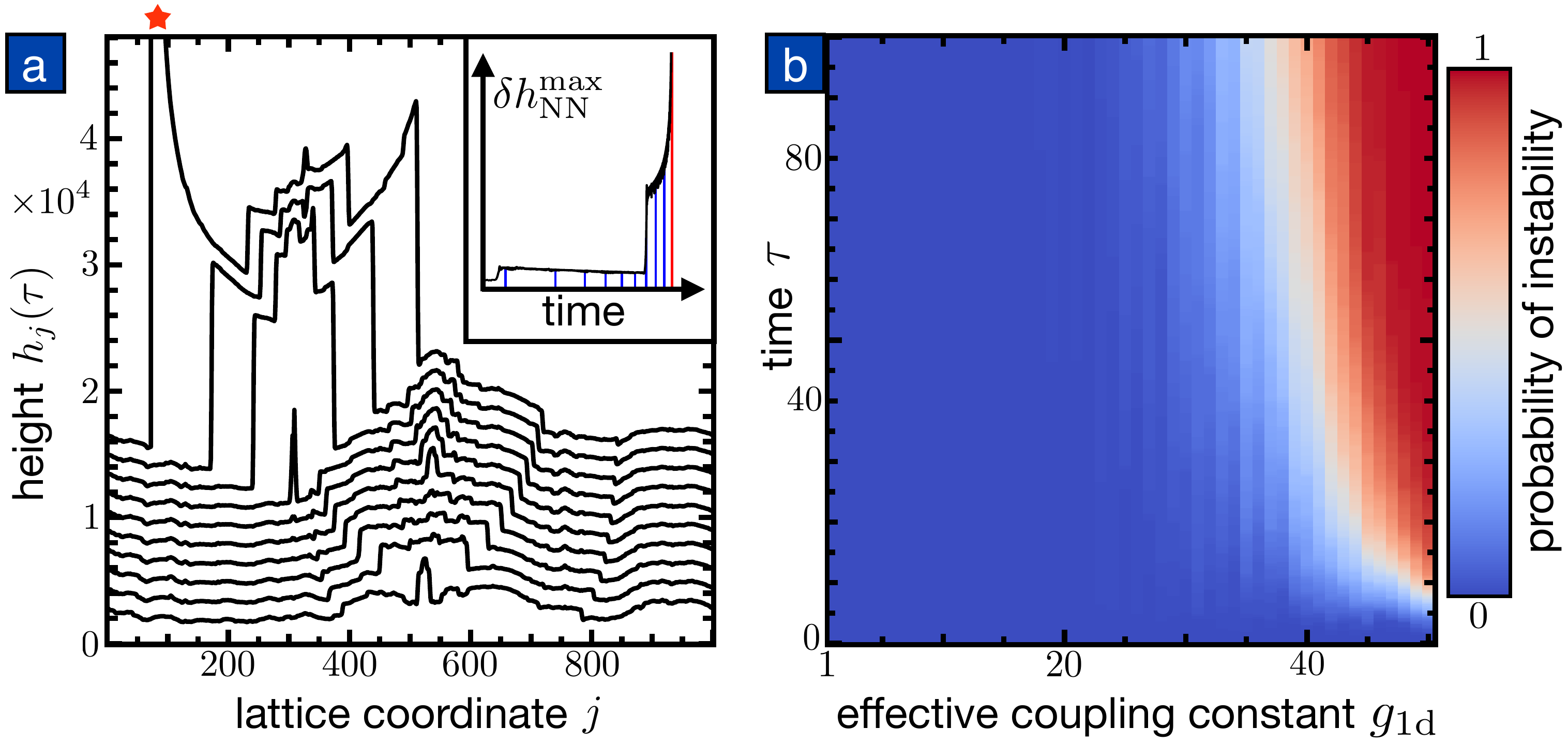}
\par\end{centering}

\protect\caption{Instabilities in the one-dimensional lattice KPZ model as given by
Eq.~(\ref{eq:KPZ-2}). (a) Typical time evolution of the height field
$h_{j}(\tau)$ for large coupling parameter $g_{\mathrm{1d}}=50$,
on a lattice with 1000 sites. We plot the height field for increasing
time from bottom to top. The curves are vertically offset for clarity.
The numerical divergence occurs at the point marked with a red star.
The selected time points approach the divergence time logarithmically,
as indicated in the inset. There, we also show the evolution of the
maximum nearest neighbor difference, $\delta h_{\mathrm{NN}}^{\mathrm{max}}$,
just before the divergence. (b) The probability of encountering an
instability up to time $\tau$, as a function of the coupling $g_{\mathrm{1d}}$.
We see that an instability is more likely to occur earlier for increasing
values of $g_{\mathrm{1d}}$. Note that the probability of instabilities
depends on the lattice size.\label{fig:1d_KPZ-2_instabilities}}
\end{figure}

To understand this important dynamical feature better, we now briefly
turn away from the full phase model and study the evolution of the
lattice KPZ model, Eq.~(\ref{eq:KPZ-2}). This serves as an approximate
description at small phase differences, so we can expect to learn
something about the onset of the growth, but not about the long-time
regime which involves large phase differences. As an example, we show
the result of a simulation of Eq.~(\ref{eq:KPZ-2}) in Fig.~\ref{fig:1d_KPZ-2_instabilities}a,
where we plot the field $h_{j}(\tau)$ for several points in time.
Clearly, even this simpler model already displays some kind of instability,
which now leads to an apparent (numerical) finite-time singularity.
It is worthwile to note that such divergences had been identified
before in numerical attempts to solve the KPZ dynamics by discretizing
it on a lattice \cite{Newman_Bray_discrete_KPZ_1996,Dasgupta_DasSarma_Instability_1997,1996_Dasgupta_DasSarma_Kim_Controlled_instability_and_multiscaling}
(see also \cite{2010_Wio_Revelli_Deza_Escudero_LaLama_Discretization-related_issues_KPZ,2008_Miranda_Reis_Numerical_Study_of_KPZ}).
In those simulations, this behavior was considered to be a numerical
artifact depending on the details of the discretization. In contrast,
in view of our phase model, the onset of the instabilities is a physical
phenomenon which merits closer inspection.

The points in time, for which the snapshots are shown in Fig.~\ref{fig:1d_KPZ-2_instabilities}a,
approach the time of the singularity logarithmically. In addition
to the normal roughening process, which we expect from the continuum
theory, we see the rapid growth of single peaks. Those can send out
shocks of large height differences, which then propagate through the
system, as can be seen in the center of Fig.~\ref{fig:1d_KPZ-2_instabilities}a.
The collision of such shocks can produce larger peaks. We commonly
observe that eventually very large shocks grow during propagation,
which leads to the singularity in the numerical evolution (marked
with a red star in the figure). In the inset, we show how the maximum
phase difference between nearest neighbors, $\delta h_{\mathrm{NN}}^{\mathrm{max}}$,
increases drastically just before the divergence. We also indicate
the points in time for which we plotted the height field. The details
of the instability development depend on the lattice size and the
coupling parameter. 

The occurrence of an instability is a random event. In Fig.~\ref{fig:1d_KPZ-2_instabilities}b,
we plot the probability of observing an instability during the evolution
up to a time $\tau$, as a function of the coupling $g_{\mathrm{1d}}$.
In principle, instabilities can occur at all coupling strengths, but
we find that for the lattice size employed here (1000 sites) they
become much less likely (happen much later) for $g_{{\rm 1d}}<40$.
To extrapolate to larger lattices, we may adopt the assumption that
the stochastic seeds for these instabilities are planted independently
in different parts of the system. In that case, the probability to
encounter a divergence within a small time interval will just scale
linearly in system size, and the present results for $N=1000$ are
therefore sufficient to predict the behavior at any $N$. 

As mentioned above, the instabilities in lattice KPZ models are considered
unphysical in the surface growth context, because they do not show
up in the continuum model, at least in one dimension \cite{Dasgupta_DasSarma_Instability_1997}.
On the contrary, our phase model, describing synchronization in discrete
oscillator lattices, is a genuine lattice model from the start. Hence,
the onset of instabilities has to be taken seriously. In the full
phase model, Eq.~(\ref{eq:Kuramoto-Sakaguchi_model}), the incipient
divergences are eventually cured by the periodicity of the coupling
functions. Instead of resulting in a finite-time singularity, they
will lead the system away from KPZ-like behavior and make it enter
a new dynamical regime. 

To find out for which parameters this happens, we have determined
numerically the probability of encountering large growth of nearest-neighbor
phase differences. We find that we can distinguish between a ``stable''
regime, where no large phase differences ($>\pi$) show up in most
simulations, and an ``unstable'' regime, where large differences
occur with a high probability. For small $S/C\ (<0.001)$, we indeed
get quantitative agreement with the results discussed above for the
lattice KPZ model, Fig.~\ref{fig:1d_KPZ-2_instabilities}b. 

\begin{figure}[t]
\centering{}\includegraphics[width=1\columnwidth]{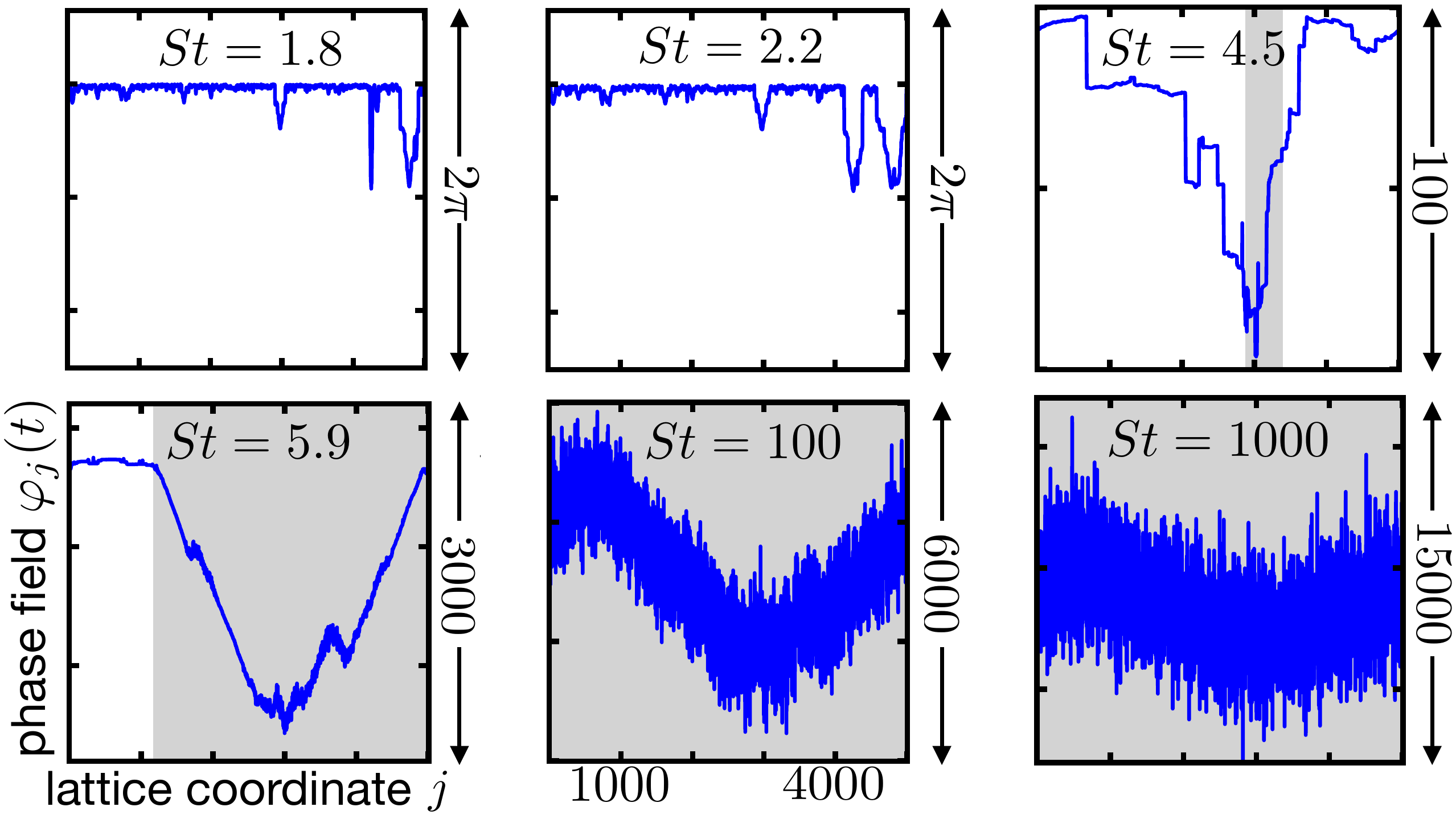}\protect\caption{Phase field time evolution in the one-dimensional phase model, Eq.~(\ref{eq:Kuramoto-Sakaguchi_model}),
in a simulation where instabilities occur. We show snapshots of the
phase field for increasing times. After a KPZ-like time evolution
with small phase differences in the beginning, we see the onset of
instabilities. This leads to the rapid development of a triangular
structure, which gets diffused on very long time scales. This phenomenon
can be explained by turbulent deterministic dynamics (in the gray
regions), see the main text. Note the very different scale of the
vertical axes in the subsequent panels. Parameters: $S/C=0.001,\ D_{\varphi}/S=1.25\times10^{-5},\ S\Delta t=10^{-4}$
(resulting in $g_{\mathrm{1d}}=50$).\label{fig:Phase_field_time_evolution}}
\end{figure}

In a single simulation in the unstable regime of the phase model,
we typically observe a time evolution such as the one depicted in
Fig.~\ref{fig:Phase_field_time_evolution}. Initially, the phase
field develops as in the corresponding KPZ lattice model. Then, a
KPZ-like instability induces large phase differences. As mentioned
above, this does not lead to a divergence. Instead, we find that huge
triangular structures develop rapidly. Afterwards, these structures
get diffused on a much longer time scale. The time evolution is reflected
in the phase spread, as shown previously in Fig.~\ref{fig:1d_phase_field_width}b
(gray lines): The development of triangular structures leads to an
explosive growth, whereas the subsequent diffusion leads to the asymptotic
scaling $w_{\varphi}\sim t^{1/2}$. 

The peculiar time evolution after the onset of the instability can
be explained by considering the deterministic phase model. For the
parameter value $S/C$ employed here, this model is (at least for
some time) turbulent for initial states with large phase differences.
In the simulations of the full model, the stochastic dynamics induces
an instability initially, which brings the phase field from a KPZ-like
state to a turbulent state locally. After this, the dynamics can be
understood deterministically. Because of the large phase differences
in the turbulent region, this part of the lattice will have a very
different phase velocity from the KPZ-like region (on average). At
the same time, the turbulent region, which is the shaded region in
the plots of Fig.~\ref{fig:Phase_field_time_evolution}, grows in
space. These two processes lead to a triangular phase field shape
covering the whole lattice. Additionally, the turbulent dynamics produces
very large phase differences, including wrap-arounds by $2\pi$. This
induces a diffusive growth of the phase field width $w_{\varphi}$
with a large diffusion coefficient. This can be seen in the red curve
of Fig.~\ref{fig:1d_phase_field_width}b. The behavior of this curve
after the rapid increase can be fitted well with $w_{\varphi}(t)=\sqrt{A+Bt}$
(blue dotted line). We checked that the diffusion coefficient $B$
from this fit can also be found in simulations of the deterministic
model with random initial conditions. The numerical value of $B$
is much larger than the noise strength $D_{\varphi}$.

Hence, we conclude that in the unstable regime of the one-dimensional
phase model, the onset of KPZ-like instabilities induces an explosive
desynchronization of the oscillators. This is followed by diffusive
growth of $w_{\varphi}(t)$. Note that there remains the large phase
field spread resulting from the desynchronization, and the large long-time
diffusion coefficient $B$, which stems from the deterministic turbulent
dynamics. All of this is relevant for small values of the parameter
$S/C$.

The physics of surface growth depends crucially on the dimensionality.
Correspondingly, we ask how the synchronization dynamics in oscillator
lattices changes when we proceed to 2D lattices, which can be implemented
in experiments and which are expected to be favorable towards synchronization. 

By using the same rescaling as above, the lattice KPZ model can be
written in dimensionless units with a single parameter $g_{\mathrm{2d}}=D\lambda^{2}/\nu^{3}$.
Interestingly, an appropriately rescaled form of the \emph{continuum}
KPZ model in 2D also contains this single parameter. That is in contrast
to the 1D case, where the rescaled continuum model did not depend
on any parameter. As a consequence, there are now different time regimes
in the growth of the surface width \cite{Nattermann_Tang_1992}. In
particular, KPZ power-law scaling $w\sim t^{\beta}$ sets in beyond
a time scale $t^{*}$ that becomes exponentially large at small couplings,
$t^{*}\sim\exp(16\pi/g_{\mathrm{2d}})$. This has to be taken into
account in numerical attempts to observe the scaling regime, as in
\cite{1990_Amar_Family_Numerical_Solution_of_KPZ_in_2d}. In finite
systems, the surface width saturates eventually, for times $(\lambda^{2}/\nu)t\gg(\lambda L/\nu)^{z}$. 

\begin{figure}[t]
\centering{}\includegraphics[width=1\columnwidth]{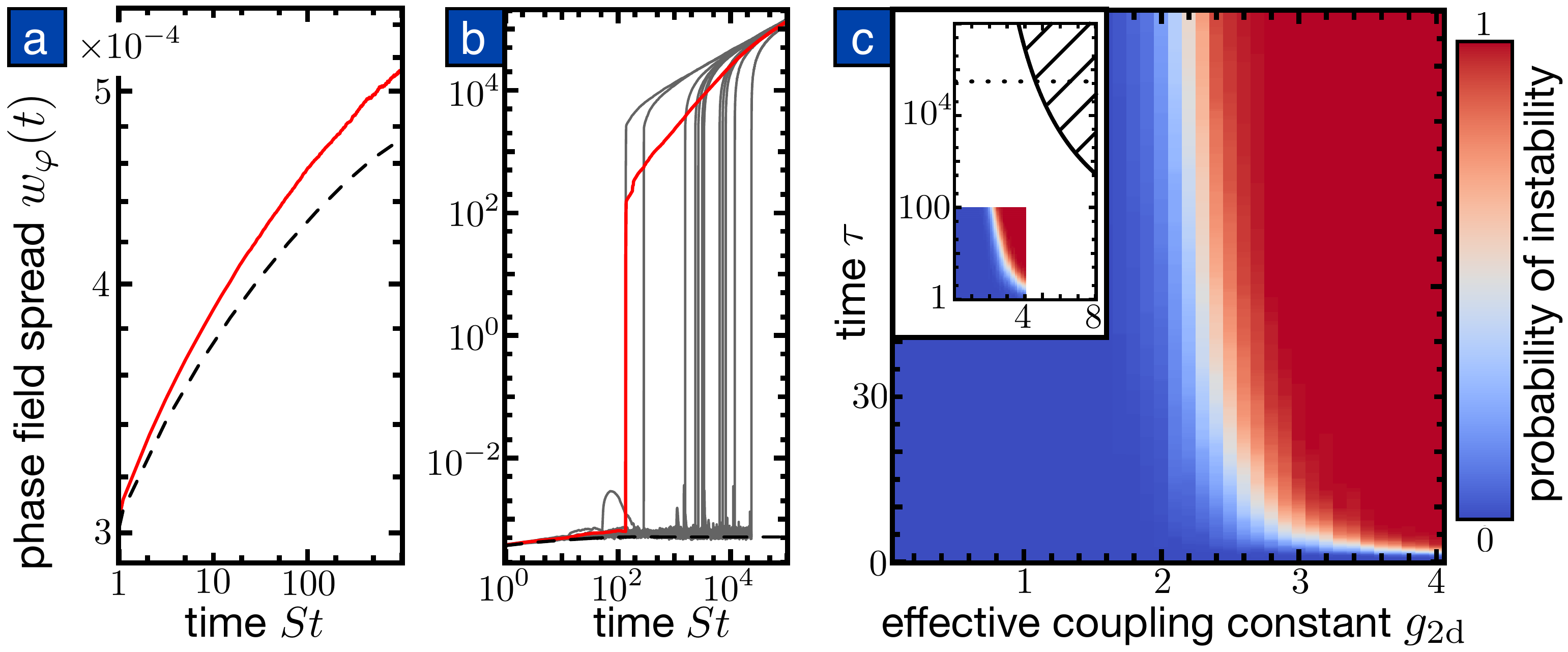}\protect\caption{Dynamics in two-dimensional models. (a) Phase model, slow logarithmic
growth of the phase spread for $g_{\mathrm{2d}}=1$ (red curve). The
data is from an average over 300 simulations with parameters $S/C=0.001,\ D_{\varphi}/S=2.5\times10^{-7},\ N=256^{2}$,
$S\Delta t=0.1$. The linear theory would lead to a slightly different
behavior, as shown by the dashed black lines in (a) and (b). (b) Same
quantity for a slightly larger coupling, $g_{\mathrm{2d}}=1.5$ (red
curve). Due to explosive instabilities (single trajectories shown
as gray lines), there is a rapid increase to much larger values than
in the linear theory. {[}Parameters: $D_{\varphi}/S=3.75\times10^{-7},\ N=64^{2}$,
otherwise like in (a){]} (c) Lattice KPZ: Probability of instability
in the lattice KPZ model, the 2D version of Eq.~(\ref{eq:KPZ-2}).
The inset shows that the power-law 2D KPZ scaling (hatched region)
would be expected at much later times than the instabilities (note
the logarithmic scaling of the time axis). This makes the scaling
unobservable also in the phase model, where the instabilities induce
a different dynamical regime.\label{fig:2d_w_sq_and_KPZ_instablities}}
\end{figure}

The lattice version of the 2D KPZ model, as obtained by extending
Eq.~(\ref{eq:KPZ-2}) to two dimensions, also develops instabilities.
Like in 1D, we study the probability of encountering such instabilities,
see Fig.~\ref{fig:2d_w_sq_and_KPZ_instablities}c. We find qualitatively
the same behavior as in 1D: The likelihood of an instability during
a time $\tau$ increases rapidly with larger $g_{\mathrm{2d}}$.

There is, however, a crucial difference with respect to the 1D situation:
we find that the instabilities occur much earlier than the (exponentially
late) onset of KPZ power-law scaling. This is illustrated in the inset
of Fig.~\ref{fig:2d_w_sq_and_KPZ_instablities}c, where the hatched
region is the KPZ scaling regime expected from the continuum theory
for infinite systems. In addition, at smaller couplings, the surface
width would saturate long before the projected onset of KPZ scaling
for any reasonable lattice sizes. As an example, the dotted line in
the inset of Fig.~\ref{fig:2d_w_sq_and_KPZ_instablities}c shows
the saturation time for a lattice of size $N=10^{6}$. Overall, we
predict that in 2D the power-law KPZ scaling regime will be irrelevant
for the synchronization dynamics of oscillator lattices.

These predictions are borne out in simulations of the full phase model,
Eq.~(\ref{eq:Kuramoto-Sakaguchi_model}), in 2D (Fig.~\ref{fig:2d_w_sq_and_KPZ_instablities}a
and b). Like in one dimension, we focus on small parameter values
of $S/C$. As long as the phase differences remain small, which is
the case for small $g_{\mathrm{2d}}=4D_{\varphi}C^{2}/S^{3}$, the
behavior is analogous to the lattice KPZ model, see Fig.~\ref{fig:2d_w_sq_and_KPZ_instablities}a.
As explained above, the exponentially large times of the KPZ power-law
regime cannot be reached before instabilities set in. Instead, the
evolution shows the behavior of the linearized KPZ equation, the so-called
Edwards-Wilkinson model \cite{Edwards_Wilkinson_1982}. This produces
a slow logarithmic growth of the surface width \cite{Edwards_Wilkinson_1982,Nattermann_Tang_1992}.
In this linear model, we can also straightforwardly take into account
the effects of the lattice discretization and the finite size of the
lattice. The resulting analytical prediction is shown as the dashed
line in Fig.~\ref{fig:2d_w_sq_and_KPZ_instablities}a, with a good
initial fit and some deviations only at later times (see also the
appendix).

In simulations of the phase model with a larger parameter $g_{\mathrm{2d}}$,
we see initially the same behavior, but followed by a rapid increase
of the phase field spread with time (see Fig.~\ref{fig:2d_w_sq_and_KPZ_instablities}b,
red curve). This can be explained by the explosive growth in single
simulations (gray lines), similar to the behavior in one dimension.
For different parameters, where the instabilities occur earlier, we
see that the phase spread approaches a diffusive square-root growth
for long times (not shown here).

Overall, we see that there is a parameter regime where lattice KPZ-like
instabilities are not relevant in 2D arrays. Then, the phase field
spreads very slowly (logarithmically) with time. According to Eq.~(\ref{eq:w_phi_sq}),
this means that the oscillators tend to synchronize quickly. However,
if instabilities show up, which is the case for larger $g_{\mathrm{2d}}=4D_{\varphi}C^{2}/S^{3}$,
we find the same explosive desynchronization as in 1D.

In conclusion, we have studied the phase dynamics of one- and two-dimensional
arrays of identical limit-cycle oscillators, described by the noisy
Kuramoto-Sakaguchi model with local coupling. We have shown that,
depending on parameters, the coupling can either enhance or hinder
the synchronization when starting from homogeneous initial conditions.
In 1D, for sufficiently small noise and at short times, one can observe
roughening of the phase field as in the Kardar-Parisi-Zhang model
of surface growth, with the corresponding universal power-law scaling.
At larger noise, or for larger times, explosive desynchronization
sets in, triggering a transition into a different dynamical regime.
We have traced back this behavior to an apparent finite-time singularity
of the approximate (KPZ-like) lattice model. This is especially relevant
for two dimensions, where it will occur before the long-term KPZ scaling
can be observed, although the initial slow logarithmic growth still
makes 2D arrays more favorable for synchronization.

With these results, we have also made more precise the connection
between phase-only models of limit-cycle oscillators and the KPZ model,
which was only established formally before \cite{Synchronization_universal_concept}.
In particular, we have shown that the lattice nature of the phase
model, Eq.~(\ref{eq:Kuramoto-Sakaguchi_model}), is important, especially
for large values of the coupling parameter $g_{\mathrm{1d,2d}}$.
The reason is that for small phase differences, we are led to a particular
lattice KPZ model, Eq.~(\ref{eq:KPZ-2}), which, however, contains
instabilities. These will destroy any resemblance between the phase
dynamics and surface growth physics. 

Our predictions will be relevant for all studies of synchronization
in locally coupled oscillator lattices, when the phase-only description
is applicable. This can be the case in optomechanical arrays (e.g.~in
extensions of the work presented in \cite{Zhang_Lipson_Synced_array_PRL}).
They may also become important for the study of driven-dissipative
condensates, described by the stochastic complex Ginzburg-Landau equation
or Gross-Pitaevskii-type equations, where a connection to the KPZ
model has been explored recently \cite{He_Diehl_2015,2016_Keeling_Sieberer_Alman_Chen_Diehl_Toner_Superfuidity_and_Phase_Correlations,2015_Ji_Gladilin_Wouters_Temporal_Coherence_of_quantum_fluids}
for the continuum case. Once these studies are extended to lattice
implementations of such models (e.g.~in optical lattices), one may
encounter the physics predicted here. 
\begin{acknowledgments}
RL and FM acknowledge support from an ERC Starting Grant and the ITN
cQOM. AM acknowledges support from National Science Foundation Grant
NSF-DMR 1303177.
\end{acknowledgments}

\section*{Appendix: Methods}

The numerical time integration of the coupled Langevin equations on
the lattice was performed with the algorithm presented in \cite{Chang_1987_numerics}.
In the following, we provide further details on the parameters employed
for the simulations whose results are shown in the figures.

For the simulations of the full phase model in one dimension in Fig.~\ref{fig:1d_phase_field_width},
we employed the following parameters. Fig.~\ref{fig:1d_phase_field_width}a:
\textbf{$S/C=0.001,\ D_{\varphi}/S=2\times10^{-6},\ S\Delta t=0.01,\ N=5\times10^{3}$},
(resulting in $g_{\mathrm{1d}}=8$). We only show a part of the phase
field. Fig.~\ref{fig:1d_phase_field_width}b: Parameters for the
upper magenta curve:\textbf{ $S/C=0.001,\ D_{\varphi}/S=2\times10^{-6},\ S\Delta t=0.01,\ N=10^{4}$}.
Lower magenta curve: \textbf{$S/C=0.001,\ D_{\varphi}/S=2.5\times10^{-7},\ S\Delta t=0.1,\ N=10^{4}$}.
For both magenta curves, the average was taken over 300 simulations.
For the red curve: \textbf{$S/C=0.001,\ D_{\varphi}/S=1.25\times10^{-5},\ S\Delta t=0.001,\ N=10^{3}$}.\textbf{
}For the green curve: \textbf{$S/C=0.1,\ D_{\varphi}/S=0.0625,\ S\Delta t=0.001,\ N=10^{3}$}.\textbf{
}The average was taken over 120 simulations.

We now turn to the simulations of the KPZ model. In general, direct
numerical simulations of this model where the scaling properties are
extracted are always performed for stable evolution. Hence, they are
done in the small-coupling regime, also for slightly different lattice
realizations with quantitatively different stability properties, see
\cite{Moser_Kertesz_Wolf_KPZ_numerics}. There, it is also found that
the parameter $g_{\mathrm{1d}}$ has an influence on the transient
dynamics in one dimension (see also \cite{1993_Forrest_Toral_Crossover_and_finite-size_effects_in_1d_KPZ})
which explains the transients that we observed in the phase model,
in Fig.~\ref{fig:1d_phase_field_width}b (magenta curves).

In Fig.~\ref{fig:1d_KPZ-2_instabilities}b, we plot the probability
of encountering instabilities in the 1D KPZ lattice model as given
by Eq.~(\ref{eq:KPZ-2}), for a wide range of the coupling parameter
$g_{\mathrm{1d}}$. The data is extracted from 300 simulations for
each value of $g_{\mathrm{1d}}=1,2,...,50$, running up to time $\tau=100$,
with a time step $\Delta\tau=10^{-4}$. The probability of instability
is just the ratio of unstable simulations. We checked that the results
for this quantity do not change at $g_{\mathrm{1d}}=50$ if we go
to a smaller time step of $\Delta\tau=10^{-5}$. A simulation was
considered unstable when the nearest-neighbor height difference at
one lattice site exceeded a large value, which was chosen to be $10^{5}$.
We used a lattice size of $N=1000$. The probability of an instability
generally increases for larger lattices. An exception are very small
lattices, where boundary effects can become important. 

Fig.~\ref{fig:2d_w_sq_and_KPZ_instablities}c shows the results for
the probability to find an unstable simulation in the 2D KPZ lattice
model. The data for the plot is from 300 simulations for each value
of $g_{\mathrm{2d}}=0.1,0.2,...,4$, on a lattice of size $N=64^{2}$
with time step $\Delta\tau=0.01$. A simulation was considered unstable
when one of the nearest neighbor height difference at one lattice
site exceeded a large value, which was chosen to be $10^{8}$. As
in 1D, the probability of instability depends on the lattice size. 

Regarding the results for the two-dimensional phase model, shown in
Fig.~\ref{fig:2d_w_sq_and_KPZ_instablities}a, we commented in the
main text on the analytical predictions from a finite-size lattice
version of the linear Edwards-Wilkinson model (dashed curve in the
figure). It can be seen that there are deviations between this curve
and the simulation of the phase model (red curve) at later times.
Further investigation shows that the two-dimensional lattice version
of the KPZ model (in analogy to Eq.~(\ref{eq:KPZ-2})) shows the
same deviations. We checked that another lattice version of KPZ (as
in \cite{Moser_Kertesz_Wolf_KPZ_numerics}) does indeed agree with
the result from the linear equation. The reason for the discrepancy
in different lattice models might be more subtle influences of the
nonlinearity, as also reported in \cite{Lam_Shin_KPZ_anomaly}.

\end{document}